\documentclass[aps,pra,groupedaddress,onecolumn,nofootinbib,12pt,a4paper]{revtex4-1}

\usepackage[german,english]{babel}

\usepackage[dvipsnames]{xcolor}
\usepackage{amsmath}
\usepackage[colorlinks=true,citecolor=MidnightBlue,linkcolor=MidnightBlue]{hyperref}
\usepackage{amsfonts}
\usepackage{amssymb}
\usepackage{mathrsfs}
\usepackage{mathtools}
\usepackage{slashed}
\usepackage{xfrac}

\usepackage{soul}
\setstcolor{red}

\usepackage{tikz-cd}
\usetikzlibrary{babel,matrix,arrows,decorations.pathmorphing}

\usepackage[T1]{fontenc}
\usepackage{times}
\usepackage{fix-cm}

\usepackage{natbib}
\usepackage{graphicx}
\usepackage[labelfont=footnotesize,font=footnotesize]{caption} %style captions
\usepackage[labelfont      =scriptsize,
            font           =scriptsize,
            singlelinecheck=off,
            justification  =centering]{subcaption}
\usepackage{booktabs} %table rules
\usepackage{tabularx}
\usepackage{tabulary}
\newcolumntype{Z}{>{\centering\arraybackslash}X} %centered column

\newcommand{\ket}[1]{\mathop{\left|#1\right\rangle}}
\newcommand{\braket}[2]{\ensuremath{\mathopen{}\left\langle{#1}|{#2}\right\rangle}}
\newcommand{\bpm}{\begin{pmatrix}}
\newcommand{\epm}{\end{pmatrix}}

\newcommand{\I}{\mathrm{i}}

\newcommand{\hilb}{\mathscr{H}}

\newcommand{\mink}{\EE^{1, 3}}

\newcommand{\Spin}{\mathrm{Spin}}
\newcommand{\SpinThree}{\mathrm{Spin(3)}}
\newcommand{\SUTWO}{\mathrm{SU}(2)}

\newcommand{\SOTHREE}{\mathrm{SO}(3)}

\newcommand{\poincare}{Poincar{\'e}\ }
\newcommand{\sutwo}{\mathfrak{su}(2)}
\newcommand{\sothree}{\mathfrak{so}(3)}
\newcommand{\Cl}{\mathrm{Cl}}
\newcommand{\dd}[1]{\mathrm{d}#1}
\newcommand{\dddd}{\mathrm{d}}

\newcommand*\diff{\mathop{}\!\mathrm{d}}
\newcommand{\ppp}{\mathbf{p}}

\newcommand{\vvv}{\mathbf{v}}

\newcommand{\pppp}{p}

\newcommand{\CC}{\mathbb{C}}
\newcommand{\EE}{\mathbb{E}}

\newcommand{\RR}{\mathbb{R}}
\newcommand{\RRR}{\mathbb{R}^3}

\newcommand{\VC}{\mathscr{V}}
\newcommand{\comment}[1]{}

\begin{document}

\title{Thomas--Wigner rotation as a holonomy for spin-$1/2$ particles}
\author{Veiko Palge}
\email[Email: ]{veiko.palge@ut.ee}
\affiliation{Laboratory of Theoretical Physics, Institute of Physics,
University of Tartu, W. Ostwaldi 1, 50411 Tartu, Estonia}

\author{Christian Pfeifer}
\email[Email: ]{christian.pfeifer@zarm.uni-bremen.de}
\affiliation{Center of Applied Space Technology and Microgravity (ZARM),
University of Bremen, Germany}

\begin{abstract}
The Thomas--Wigner rotation (TWR) results from the fact that a combination of
boosts leads to a non-trivial rotation of a physical system. Its origin lies in
the structure of the Lorentz group. In this article we discuss the idea that
the TWR can be understood in the geometric manner, being caused by the
non-trivially curved relativistic momentum space, i.e.\ the mass shell, seen as
a Riemannian manifold. We show explicitly how the TWR for a massive spin-$1/2$
particle can be calculated as a holonomy of the mass shell. To reach this
conclusion we recall how to construct the spin bundle over the mass shell
manifold.
\end{abstract}

%\pacs{}
\maketitle

\tableofcontents

\section{Introduction}

The Thomas--Wigner rotation (TWR) is a fascinating effect of special
relativity, which originates in the fact that a combination of boosts results
in a non-trivial rotation of a physical system \cite{ThomasOriginal,
WignerOrig, Malykin_2006, ODonnell:2011ivw, Kholmetskii_2020}. If a
non-relativistic system moving with velocity $\vvv_1$ is boosted by velocity
$\vvv_2$, the resulting velocity is given by the familiar law of addition of
velocities $\vvv = \vvv_1+\vvv_2$. In special relativity this is not the case.
Two successive non-collinear boosts lead to a boost \emph{and rotation}. This
phenomenon is called TWR and it originates in the structure of the Lorentz
group, which encodes the fundamental symmetries of special relativity and
Minkowski spacetime.

An alternative approach to understand the TWR is a geometric one. The TWR can
be thought of as being caused by the non-trivially \emph{curved} relativistic
momentum space of massive particles, the mass shell, seen as a Riemannian
manifold. The goal of this paper is to describe TWR in the context of
relativistic quantum theory using the geometric approach. We focus on the free
massive spin-$1/2$ particle. Whereas the standard approach in quantum theory is
to use group theory and the Hilbert space formalism, the advantage of the
geometric approach lies in its highly intuitive conceptualization of TWR. It
explains the non-intuitive character of TWR in terms of the fact that the
momentum space is a curved Riemannian manifold. This is in contrast to the
non-relativistic momentum space where the familiar law for addition of
velocities means that it is a flat Euclidean space $\RRR$.

The geometric approach to TWR dates back to almost the birth of special
relativity, see for instance \cite{rhodes_relativistic_2004} where the idea of
a hyperbolic velocity space is tracked to the articles published between 1910
and 1919. In more recent literature, the hyperbolic velocity space is described
and derived from somewhat different starting points \cite{aravind_1997,
criado_2001, rhodes_relativistic_2004}, where the last reference in particular
has inspired the present paper.

We build on prior work in the geometric approach. We start with the common
underlying idea that the relativistic momentum, or equivalently, velocity space
is a curved Riemannian manifold and use the language of differential geometry
to develop the notion that the TWR is nothing but a holonomy of the
relativistic momentum space. Holonomy is the idea that when a vector (or
spinor) is parallel transported along a closed curve, then the initial and the
final vector (or spinor) need not necessarily coincide because the manifold is
curved. The transformation between the initial and the final, parallel
transported vector is described by the holonomy matrix. This means that TWR
belongs to the larger family of classical and quantum physical phenomena which
can be described as holonomies, such as Berry's phase, the Aharonov-Bohm effect
or even classical effects like the Foucault pendulum \cite{lyre_2014,
CRIADO2009923}.

The article is divided into two parts. In the first part (starting with section
\ref{sec:twr-standard-classical-quantum-1}) we review the TWR in classical and
quantum physics. Those familiar with this background may want to skip ahead to
the second part (starting with section
\ref{sec:twr-geometry-classical-quantum-1}) where we present the main results
of the article. We start by providing an intuitive picture which explains TWR
in geometric terms. This is followed by the gradual formalization of the
intuitive picture. We then describe the intrinsic geometry of the mass
hyperbola using the standard differential geometric language of fiber bundles,
connection and curvature. After that we add the quantum spin field to the
picture and show how the connection and curvature can be induced for the spinor
bundle. Finally we explicitly calculate the holonomy matrix for the Thomas
precession of the spin-$1/2$ particle. As a result we also reproduce the
holonomy angle that coincides with the result obtained in
\cite{rhodes_relativistic_2004}.

We will use the following notational conventions. Throughout we denote four
vectors by normal font with Greek indices $\mu, \nu$ running from $0$ to $3$,
where $x^0$ is the time component. Latin indices $\{ 1, 2, 3 \}$ which run over
spatial coordinates and spatial vectors $\mathbf{x}$ or $1$-forms $\mathbf{p}$
are boldfaced. The Minkowski space with Minkowski metric will be denoted by
$\mink$ and described in canonical Cartesian coordinates in which the Minkowski
metric $\eta$ has the form $\eta = \textrm{diag}(1,-1,-1,-1)$. The units are
natural, $\hbar = c = 1$. The invariant four momentum of a particle is denoted
by the $1$-form $P$, which in local Cartesian coordinates can be expanded as
\begin{align}
    P = p_\mu dx^\mu
\end{align}
so the four momentum of a particle with mass $m$ is given by $p_\mu = (p_0,
\ppp)$ with norm $\eta^{\mu\nu} p_\mu p_\nu = p^\mu p_\mu = (p_0)^2 - \ppp^2 =
m^2$ where $\pppp_0 = \sqrt{m^2 + \ppp^2} =: E(\ppp)$.

\section{Thomas--Wigner rotation in classical and quantum physics}
\label{sec:twr-standard-classical-quantum-1}

In this section we will give a quick overview of the TWR in classical special
relativistic physics and in relativistic quantum mechanics. We start by
discussing how successive boosts act on classical point particle momenta. After
this we examine their action on quantum mechanical spins.

There is no geometry involved at this point. Readers familiar with the standard
account of TWR can skip ahead to section
\ref{sec:twr-geometry-classical-quantum-1} where we discuss the geometric
approach to TWR in terms of the curved relativistic momentum space. Our
approach complements and builds upon earlier treatments of TWR using projective
geometry \cite{rhodes_relativistic_2004}.

\subsection{Boosts of relativistic momenta and TWR}
\label{sec:boosts-relativistic-momenta-twr-1}

\begin{figure}[htb]
\includegraphics[width=0.25\textwidth]{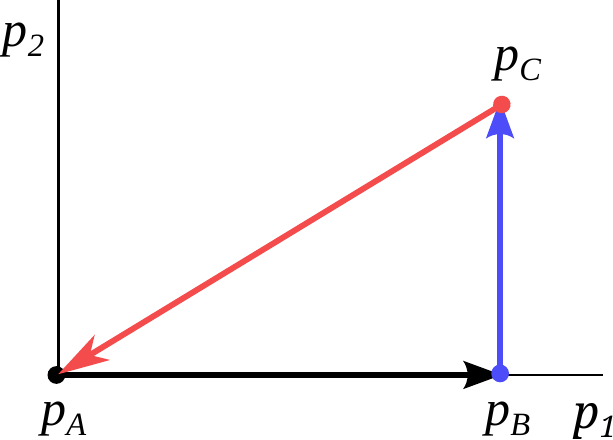}
\caption{\label{fig:boost-sequence-1}
Sequence of three boosts in the $2D$ plane of the four momentum space. The
energy axis $p_0$ is perpendicular to the plane and component $p_3 = 0$ of the
four momentum is suppressed. The plane consists of two spatial components $\ppp
= (p_1, p_2)$ where each point corresponds to a system with momentum $\ppp$.
The first boost (black arrow) with velocity $\vvv_1$ along the $p_1$ axis takes
a system from rest momentum $\ppp_A = (0, 0)$ to $\ppp_B = (p_B, 0)$. The
second boost (blue arrow) with velocity $\vvv_2$ is along the $p_2$ axis to
$\ppp_C$, and the third (red arrow) with velocity $\vvv_{12}$ brings the system
back to rest. As a result, the $2D$ plane undergoes TWR by $R(\vvv_1, \vvv_2)$
about the $p_0$ axis.}
\end{figure}

Suppose a body with mass $m$ is boosted from rest by velocity $\vvv_1$. We
assume the boost is rotation free, i.e.\ pure, then momentum undergoes the
following transformation,
\begin{align}\label{eq:boost-1}
L(\vvv_1) p_A = p_B
\end{align}
where $p_A = (m, 0, 0, 0)$ is the four momentum of the system at rest and
$L(\vvv_1)$ is a pure boost that maps the rest momentum to $p_B$. A second pure
boost with velocity $\vvv_2$ with respect to the frame with velocity $\vvv_1$,
maps $p_B$ to $p_C$
\begin{align}
L(\vvv_2) L(\vvv_1) p_A = L(\vvv_2) p_B = p_C\,.
\end{align}
If the velocity $\vvv_2$ is along the \emph{same} direction, then the resulting
velocity $v_{12}$ of the final frame $C$ with respect to the first frame $A$ is
given by the familiar formula for addition of relativistic velocities,
\begin{align}
v_{12} = \frac{v_1 + v_2}{1 + v_1 v_2},
\end{align}
where $v_i = |\vvv_i|$. However, if the second boost is \emph{not} in the same
direction but along a different direction that makes an angle $\theta$ relative
to the first boost, then the velocity addition is more involved
\begin{align}
\mathbf{v}_{12} = \frac{1}{1+\vvv_1 \cdot \vvv_2}\left( \left(1 +
\frac{\gamma(\vvv_1)}{1+\gamma(\vvv_1)} \vvv_1 \cdot \vvv_2 \right)\vvv_1 +
\frac{1}{\gamma(\vvv_1)} \vvv_2 \right) \,,
\end{align}
where $\gamma(\vvv)^{-1} = \sqrt{1 - \vvv^2}$. Importantly, the resulting
momentum $p_C$ is not the momentum that one would get with a single boost from
rest by composite velocity $\vvv_{12}$. The final momentum generally is also
additionally subject to a \emph{rotation} $R(\vvv_1, \vvv_2)$ called the
\emph{Thomas--Wigner rotation}. Formally, this means
\begin{align}\label{eq:pure-boosts-to-rotation-1}
L(\vvv_2) L(\vvv_1) = L(\vvv_{12}) R(\vvv_1, \vvv_2),
\end{align}
where $R(\vvv_1, \vvv_2)$ is a rotation by angle $\alpha$ that depends on
velocities $\vvv_1$ and $\vvv_2$. Using Eq.\
(\ref{eq:pure-boosts-to-rotation-1}) we can express the rotation as a sequence
of boosts,
\begin{align}\label{eq:twr-by-r-v1-v2}
R(\vvv_1, \vvv_2) = L(\vvv_{12})^{-1} L(\vvv_2) L(\vvv_1).
\end{align}
This is the standard expression of the TWR as a result of a sequence of three
boosts: first boosting the system from rest along arbitrary directions $\vvv_1$
and $\vvv_2$, and then bringing it back to rest by the third boost as shown in
Fig.\ \ref{fig:boost-sequence-1}.
For non-relativistic velocities, the angle $\alpha$ by which the TWR rotates
the system is negligible. In ultra-relativistic scenarios it can approach $180$
degrees, depending on the geometry of the boost situation, i.e.\ the angle
between the velocities $\vvv_1$ and $\vvv_2$, and the magnitudes of the
velocities.

In summary, the TWR always occurs when boosts are \emph{non-collinear}. This
can be generalized to an arbitrary number of boosts where at least two are
non-collinear. Group theoretically, the reason is that the subset of boosts in
the Lorentz group does not form a subgroup.\footnote{This is in contrast to the
non-relativistic situation: boosts in the Galilei group do form a subgroup.}
Instead, a combination of two boosts results in a boost \emph{and} a rotation
as expressed in Eq.\ \eqref{eq:pure-boosts-to-rotation-1}.

TWR gives rise to a plethora of effects in different branches of physics. It
can be measured for satellites moving around the Earth, it manifests as a
correction term for the non-relativistic Hamiltonian of the hydrogen atom and
it must be taken into account when calculating scattering cross sections in
quantum field theory. It is also the reason why the behavior of quantum
entanglement in relativity is significantly different from its non-relativistic
counterpart \cite{peres_quantum_2002, gingrich_quantum_2002,
caban_2009_strange, friis_2010_relativistic, palge_2015_werner,
palge_2019_maps, barr_2023_bell}. Because it is ultimately a feature of the
structure of relativistic spacetime,\footnote{This is sometimes expressed by
saying that the TWR is a \emph{kinematic} effect.} the TWR is independent of
the dynamics that caused the boost \cite{rhodes_relativistic_2004}.

\subsection{Quantum mechanical spins and TWR}\label{sec:qm-spins-and-twr-1}

In this section we will summarize the standard treatment of massive spin-$1/2$
particles in the Hilbert space formalism in relativistic quantum mechanics.

In the Hilbert space theory of quantum systems, particles with spin are
described as representation spaces of the relevant symmetry group. Free massive
spin-$1/2$ particles can be described by two different but equivalent theories,
the first using the unitary irreducible representations of the \poincare group
and the second the finite-dimensional representations of the Lorentz
group.\footnote{See \cite{polyzou_spin_2012} for a good overview, including a
detailed explanation of how the two approaches are related.} We will work in
the first approach which relies on the Wigner representation (also called the
Wigner--Bargmann or the spin basis), and is presented in the references
\cite{bogolubov_introduction_1975, sexl_relativity_2001, caban_spin_2013}. Here
the single particle states are given by the unitary representations of the
\poincare group which are labelled by mass $m > 0$ and the intrinsic spin $s$,
where the latter takes both integer and half-integer values. The
representations are realized in the space $\hilb_{m, s}^+ = \bigoplus^{2s + 1}
L^2(\VC^+_m)$ of square integrable functions on the forward mass hyperboloid
$\VC^+_m = \{ p_\mu \in \mink \;|\; \eta^{\mu\nu} p_\mu p_\nu = m^2, p_0 > 0
\}$ where the scalar product is defined as
\begin{align}
\braket{\phi}{\psi}
&= \sum_{\sigma = 1}^{2 s + 1} \int \dddd\mu(\ppp)\, \phi^*_{\sigma}(\ppp)
\psi_{\sigma}(\ppp)
\end{align}
with $\dddd\mu(\ppp) = [2E(\ppp)]^{-1}\dddd^3\ppp$ being the Lorentz invariant
integration measure and $\phi_\sigma(\ppp), \psi_\sigma(\ppp)$ elements of
$L^2(\VC^+_m)$. The state space $\hilb_{m, 1/2}^+ $ of single spin-$1/2$
particle with mass $m$ is given by $L^2(\VC^+_m) \otimes \CC^2$. Using basis
states which are labeled by the three momentum $\ppp$ and spin $\sigma$, a
generic state can be written as
\begin{align}
\ket{\psi} = \sum_{\sigma} \int \dddd\mu(\ppp)\, \psi_\sigma(\ppp) \ket{\ppp,
\sigma}.
\end{align}
The general Lorentz transformation $\Lambda$ acts on the basis element as
follows,
\begin{align}\label{eq:lorentz-unitary-rqm-1}
U(\Lambda) \ket{\ppp, \sigma} = \sum_{\lambda} \ket{\Lambda\ppp, \lambda}
D_{\lambda\sigma}[W(\Lambda, \ppp)],
\end{align}
where we write $\Lambda \ppp$ for the spatial part of the vector $\Lambda p$,
with $p = (E(\ppp), \ppp)$, and $W(\Lambda, \ppp)$ is the Wigner rotation,
\begin{align}\label{eq:standard-wigner-rotation-w-l-lambda-1}
W(\Lambda, \ppp) = L^{-1}(\Lambda\ppp) \Lambda L(\ppp)
\end{align}
which leaves $p_0$ invariant. Note that $W(\Lambda, \ppp)$ exhibits the same
form as $R(\vvv_1, \vvv_2)$ in \eqref{eq:twr-by-r-v1-v2} since $L(\ppp)$ is the
boost that maps the rest momentum to $\ppp$, $\Lambda$ performs an arbitrary
boost and $L^{-1}(\Lambda\ppp)$ maps the system back to rest. For massive
particles, $W$ is an $\SOTHREE$ rotation and $D[W(\Lambda, \ppp)]$ the
corresponding representation. The latter is an element of $\SUTWO$ for
spin-$1/2$ particles and it can be generally written as
\begin{align}
D(\alpha) = \exp\left(-\I \,\alpha\, \hat{\boldsymbol{n}} \cdot
{\boldsymbol\sigma} / 2 \right),
\end{align}
where $\alpha$ is the Wigner rotation angle and the three unit vector
$\hat{\boldsymbol{n}}$ defines the rotation axis. This rotation matrix can be
parameterized in terms of momenta and rapidities \cite{halpern_special_1968}.

We can now relate the action of boost operator $U(\Lambda)$ in
Eq.~(\ref{eq:lorentz-unitary-rqm-1}) to the two-boost scenario described above
in Fig.~\ref{fig:boost-sequence-1}. The label $\ppp$ in the quantum state
$\ket{\ppp, \sigma}$ of Eq.~(\ref{eq:lorentz-unitary-rqm-1}) refers to a system
moving with velocity $\vvv_1 = \ppp / E(\ppp)$ after the first boost. This
corresponds to $\ppp_A$ in Fig.~\ref{fig:boost-sequence-1}. Boost $\Lambda$ in
corresponds to the second boost by velocity $\vvv_2$ in
Fig.~\ref{fig:boost-sequence-1}. The TWR $W(\Lambda, \ppp)$ is the rotation
that the quantum system undergoes as a result of boost $\Lambda$ when its state
of motion changes from $\ket{\ppp, \sigma}$ at point $\ppp_A$ to state
$U(\Lambda)\ket{\ppp, \sigma}$ at point $\ppp_B$ in
Fig.~\ref{fig:boost-sequence-1}.

But why do two non-collinear boosts lead to a rotation? In the next section we
will see that interpreting the boost sequence in
Fig.~\ref{fig:boost-sequence-1} in the geometric context provides a natural
explanation of the phenomenon.

\section{The geometry of curved momentum space and TWR}
\label{sec:twr-geometry-classical-quantum-1}

In this section we embark on the geometric study of TWR. We begin by discussing
the intuitive picture of boosts from the perspective of the curved momentum
space. Section \ref{sec:geometry-of-curved-momentum-space-1} describes the
geometry of the mass hyperboloid in terms of standard differential geometry.
Thereafter we discuss quantum spins in terms of spinor bundles over the mass
hyperboloid. We give an explicit derivation of the bundle connection in section
\ref{sec:spinors-on-mass-hyperboloid-1} and \ref{sec:spinor-connection-1}.
Finally in section \ref{sec:TWRHolo} we conceptualize TWR as the holonomy of
the curved momentum space.

\subsection{Prelude: the intuitive picture}
\label{sec:prelude-twr-in-projective-geometry-1}

The account of how a sequence of non-collinear Lorentz boosts results in a
rotation, and why they behave differently from the more familiar Galilei boosts
discussed in Section \ref{sec:twr-standard-classical-quantum-1}, relies on
\emph{group theoretic} properties of Lorentz boosts. However, an alternative
perspective on TWR gives a \emph{geometric} explanation for its appearance. In
terms of projective geometry this is discussed in
\cite{rhodes_relativistic_2004}, and in \cite{sexl_relativity_2001} the authors
consider the curved velocity space.

\begin{figure}[htb]
\includegraphics[width=0.25\textwidth]{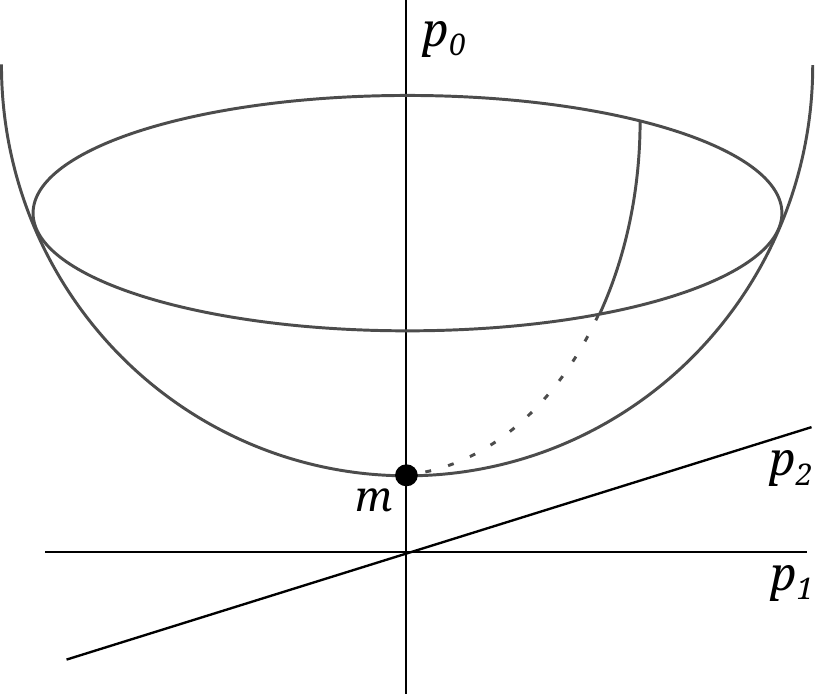}
\caption{\label{fig:velocity-hyperboloid-1}The $2D$ relativistic mass
hyperboloid. Component $p_3 = 0$ is not shown. Black dot indicates rest
momentum at $p_A = (m, 0, 0,0)$ where the hyperboloid intersects the $p_0$
axis.}
\end{figure}

Our aim is to discuss the TWR as being caused by curved momentum space. This
picture is particularly suitable since quantum mechanical spinors in momentum
space representation can be understood precisely in terms of a spinor bundle
over the relativistic curved momentum space. The latter is also key to
understanding the reason why Lorentz boosts behave differently from Galilei
boosts. While the non-relativistic momentum is given by $\RR^3$, which a flat
space, the relativistic momentum space is given by a curved manifold---the mass
hyperboloid, shown in the $2D$ case in Fig.~\ref{fig:velocity-hyperboloid-1}.

Points on the hyperboloid correspond to physically viable momenta, satisfying
the relativistic dispersion relation $\eta^{\mu\nu}p_\mu p_\nu = m^2$. For
instance, we can consider a physical system (an inertial frame or a spin of a
particle or an observer) at point ${p_A = (m, 0, 0,0)}$ on the hyperboloid,
which corresponds to the system being at rest. There is a natural way to think
of boosts of the physical system in the curved setting. A \emph{pure boost}
corresponds to parallel transporting the system along a geodesic from the
initial to the final state of motion. For example, boosting a system from
momentum $p_A$ to $p_B$ means the system is parallel transported along a
geodesic from point $p_A$ to $p_B$ on the hyperboloid.

\begin{figure}[htb]
\includegraphics[width=0.25\textwidth]{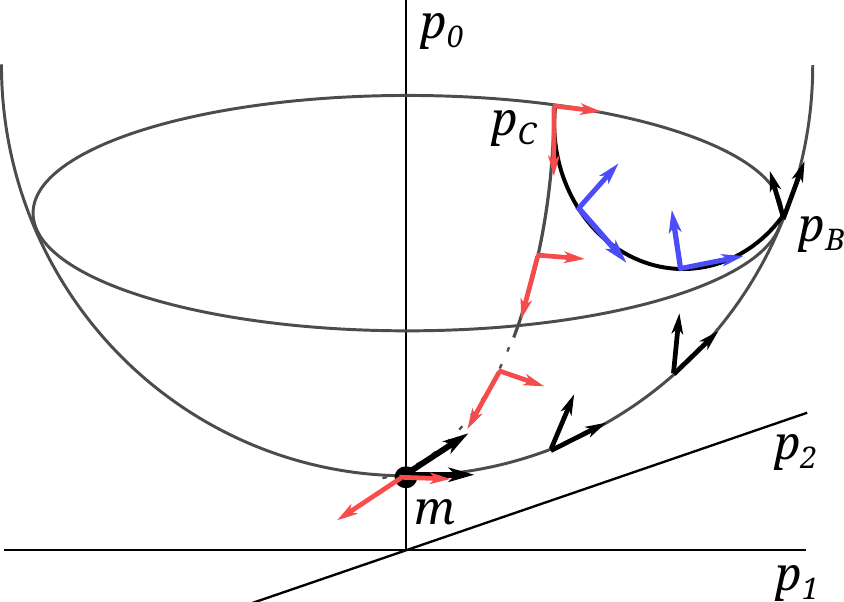}
\caption{\label{fig:boost-sequence-hyperboloid-1}Sequence of three boosts on
the $2D$ mass hyperboloid. Black dot indicates rest momentum at $p_A = (m, 0,
0,0)$ where the hyperboloid intersects the $p_0$ axis. The first (black) boost
is along the geodesic from rest to point $p_B$, the second (blue) along the
geodesic from point $p_B$ to $p_C$. The third (red) brings the system back to
rest. The resulting final frame (red) at the origin is Thomas--Wigner rotated
relative to the initial (black) frame.}
\end{figure}

Using the hyperboloid, let us visualize the action of the boost sequence
previously shown in Fig.~\ref{fig:boost-sequence-1}. For the visualization, we
consider boosting a $2D$ frame $F(p)$, see
Fig.~\ref{fig:boost-sequence-hyperboloid-1}. Without loss of generality, let us
orient this $2D$ frame initially, when it is at rest at $p_A= (m, 0, 0, 0)$,
along the $p_1$ and $p_2$ axis. The first boost $L(\vvv_1)$ transports the
frame along a geodesic of the hyperboloid (shown black) to $p_B$. There is no
rotation as a result of the boost, the frame remains oriented in the original
direction. Now, applying the second boost $L(\vvv_2)$ along the $p_2$ axis
means the frame (shown blue) travels along a geodesic which does not intersect
with the origin. Parallel transporting the frame along that geodesic to point
$p_C$ means the frame is \emph{rotated} relative to the frame at the origin. We
can see this clearly when we boost the frame (shown red) back to the origin
$p_A$ along a geodesic. This geodesic \emph{does} intersect the origin, hence
there is \emph{no rotation} involved when the frame is parallel transported
along that geodesic.
The final (red) frame shown in Fig.~\ref{fig:boost-sequence-hyperboloid-1} is
TWR rotated relative to the initial (black) frame.

This example demonstrates the power of the geometric picture. Since boosting is
conceptualized as parallel transport between points on the hyperboloid, one can
easily see why the final frame is rotated relative the initial one. Other
scenarios can be analyzed in a similar manner. In general, one recognizes that
whether or not a frame will be rotated as a result of a series of boosts
depends on the path that the frame follows on the hyperboloid. For instance,
all geodesics that intersect the origin do not rotate the frame. Rotation
occurs only when the system is boosted along a geodesic that does not intersect
the origin \cite{rhodes_relativistic_2004}. Hence parallel transport along a
generic path on the hyperboloid will give rise to rotation.

With this intuitive picture at hand, we will next give a formal, differential
geometric description of the mass hyperboloid.

\subsection{The geometry of the mass hyperboloid}
\label{sec:geometry-of-curved-momentum-space-1}

In the previous section we described the momentum space as a hyperboloid and
argued heuristically that boosting a system means parallel transporting it
along a particular trajectory. We now turn to the formal, differential
geometric description of the relativistic curved momentum space. The state
space of a spin-$1/2$ particle as a field over this manifold will be introduced
below in section \ref{sec:spinors-on-mass-hyperboloid-1} and the TWR as
holonomy in section \ref{sec:TWRHolo}.

We start with Minkowski spacetime $\mink$ and note that as a manifold, this
space comes equipped with a tangent $T_x \mink$ and cotangent space $T_x^*
\mink$ at each point $x \in \mink$. The mass hyperboloid $\VC^+_m$ is defined
as a subset of the cotangent space,
\begin{align}
\VC^+_{m,x} = \{ P \in T^*_x \mink \; | \;
\eta(P, P) = \eta^{\mu\nu} p_\mu p_\nu = m^2, p_0 > 0 \} \subset  T^*_x \mink\,.
\end{align}
Since the Minkowski space is flat, we can identify the mass hyperboloids at
different points of spacetime and omit the subscript $x$ that labels the base
point, $\VC^+_{m,x} = \VC^+_{m}$. As a result, we speak of a single momentum
hyperboloid in which the particle moves.\footnote{In general, this cannot be
done in curved spacetimes since there is no canonical way to identify
hyperboloids at neighboring points.}

To describe the intrinsic geometry of the hyperboloid, we consider $\VC^+_m$ as
the image of a mapping of spherical polar coordinates $z = (\rho, \theta,
\phi)$, where $\rho \in \left[0, \infty\right), \theta \in \left[0, \pi\right),
\phi \in \left[0, 2\pi\right)$, into the cotangent space $T_x^* \mink$,
\begin{align}\label{eq:hyppara}
f: (\rho, \theta,\phi) \mapsto \left(\sqrt{m^2+\rho^2},\rho \sin\theta\cos\phi, \rho\sin\theta\sin\phi,\rho\cos\theta \right) = p(\rho, \theta,\phi)\,,
\end{align}
where
\begin{align}
\rho = |\ppp| = \sqrt{\ppp \cdot \ppp}\, ,\quad
\tan\theta = \frac{\sqrt{p_1^2+p_2^2}}{p_3}\, , \quad
\tan\phi = \frac{p_2}{p_1}.
\end{align}
For $\theta= \pi/2$ this parameterization of the mass hyperboloid encodes motion
of a particle in the spatial $1-2$ plane. The energy $E := E(\rho)$ of the
particle is identified as
\begin{align}
E(\rho) = \sqrt{m^2 + \rho^2}\,,
\end{align}
and the relativistic $\gamma(v) = 1 / \sqrt{1 - v^2}$ factor for a particle
with velocity $v$ in this language is given by
\begin{align}\label{eq:gamma}
\gamma(v) = \frac{E}{m} = \frac{ \sqrt{m^2 + \rho^2}}{m}\,\, \Leftrightarrow \,\,
            \rho = \frac{m v}{\sqrt{1 - v^2}}\,.
\end{align}

We next induce a metric tensor $g$ on the hyperboloid via pullback of the
Minkowski inner product from $T^*_x\mink$ to $\VC^+_m$
\begin{align} \label{eq:Hmet}
g = -\left( \tfrac{m^2}{E^2}     \dd \rho  \otimes \dd \rho
  + \rho^2               \dd\theta \otimes \dd\theta
  + \rho^2 \sin^2 \theta \dd\phi \otimes \dd\phi\right) \, .
\end{align}
This metric is negative definite since we started from the Minkowski metric of
signature~$(+,-,-,-)$. It defines the intrinsic geometry of $\VC^+_m$ via the
Levi-Civita connection which determines the parallel transport of vectors, and
whose non-vanishing Christoffel symbols $\Gamma^i{}_{jk} = \frac{1}{2}g^{ip}
\left(\partial_j g_{pk} + \partial_k g_{pj} - \partial_p g_{jk}\right)$ are
given as follows,
\begin{align}\label{eq:christoffelmass}
\Gamma^{\rho}{}_{\rho\rho} &= - \tfrac{\rho}{m^2 + \rho^2}\,, & \Gamma^{\rho}{}_{\theta\theta} \sin^2\theta &= \Gamma^{\rho}{}_{\phi\phi} = - \tfrac{\rho}{m^2}(m^2 + \rho^2)\sin^2\theta\,,\\
\Gamma^{\theta}{}_{\rho\theta} &= \Gamma^{\theta}{}_{\theta\rho} = \tfrac{1}{\rho}\,, & \Gamma^\theta{}_{\phi\phi} &= - \cos\theta\sin\theta\,,\\
\Gamma^{\phi}{}_{\rho\phi} &= \Gamma^{\phi}{}_{\phi\rho} = \tfrac{1}{\rho}\,, & \Gamma^{\phi}{}_{\theta\theta} &= \cot\theta \, .
\end{align}
The Riemann curvature tensor of the hyperboloid can be also easily evaluated,
$R^p{}_{qij} = \partial_i \Gamma^p{}_{qj} - \partial_j \Gamma^p{}_{qi} +
\Gamma^p{}_{si}\Gamma^s{}_{qj} - \Gamma^p{}_{sj}\Gamma^s{}_{qi}$, and its Ricci
scalar is constant $R = g^{qj}R^p{}_{qpj} = 6/m^2$. The latter is not
surprising since the hyperboloid is a maximally symmetric space with constant
curvature.

In order to describe the parallel transport of spinors later in sections
\ref{sec:spinors-on-mass-hyperboloid-1} -- \ref{sec:TWRHolo}, we need another
geometric ingredient: the spin connection coefficients induced by the
Levi-Civita connection. Loosely speaking, we need to express the information
encoded in the Levi-Civita connection as a collection of $1$-forms
$\omega^A{}_B$ on the curved momentum manifold $\VC^+_m$,
\begin{align}
\omega^A{}_B = \sum_{i \in \{\rho, \phi, \theta\}}\omega^A{}_{B i} \diff z^i\, ,
\end{align}
whose components can be computed using the relation
\begin{align}\label{eq:connection-general-1}
\omega^A{}_{B i} = e^A{}_k e_B{}^j{} \Gamma^k{}_{ij}
+ e^A{}_k \partial_i e_B{}^k{}\,,
\end{align}
where $e^A{}_i$ are components of an orthonormal coframe $\Theta^A = e^A{}_i
\diff z^i$ of the momentum space metric~$g$. In other words, we write the
Christoffel symbols in an orthonormal frame basis $\theta^A$, while reserving
the index $i$ for the coordinate basis. The momentum space $1$-forms
$\omega^A{}_B$ are the components of a $\SOTHREE$ connection $1$-form $\omega$
on $\VC^+_{m}$, which we will later on map to the $\SUTWO$ spinor connection
over $\VC^+_{m}$.

Using the following frame and coframe of the metric \eqref{eq:Hmet},
\begin{align}\label{eq:frame}
    e_1 = \frac{E}{m}\, \partial_\rho,\;\;
    e_2 = \frac{1}{\rho} \partial_\theta,\;\;
    e_3 = \frac{1}{\rho \sin\theta} \partial_\phi,
    \;\;\quad
    \Theta^1 = \frac{m}{E} \diff\rho,\;\;
    \Theta^2 = \rho \diff\theta,\;\;
    \Theta^3 = \rho \sin\theta \diff\phi,
\end{align}
we can display the coefficients $\omega^A{}_B$ as the matrix
\begin{align}\label{eq:local-connection-form-mass-shell-1}
\omega &=
\begin{bmatrix}
0    &   -\frac{E}{m} \diff\theta &    -\sin\theta\, \frac{E}{m} \diff\phi     \\
\frac{E}{m} \diff\theta     &    0     &        -\cos\theta \diff\phi  \\
\sin\theta\, \frac{E}{m} \diff\phi     &   \cos\theta \diff\phi    &     0
\end{bmatrix}.
\end{align}
Similarly, we can express the curvature of the Levi-Civita connection as
a collection of $2$-forms~$\Omega^A{}_B$
\begin{align}
    \Omega^A{}_B =  \frac{1}{2}R^A{}_{Bij} dz^i \wedge dz^j = \frac{1}{4} R^A{}_{Bij} (dz^i \otimes dz^j - dz^j \otimes dz^i ) =\frac{1}{2} R^A{}_{Bij} dz^i \otimes dz^j  \,,
\end{align}
where $R^A{}_{Bij} = R^a{}_{bij} e^A{}_a e^b{}_B$ are the components of the
Riemann curvature tensor of the metric $g$ partially expressed in the
orthonormal frame and coframe $e$ and $\Theta$. The momentum-space $2$-forms
$\Omega^A{}_B $ are the components of the $\SOTHREE$ curvature $2$-form of the
over $\VC^+_{m}$. For the frame \eqref{eq:frame} we obtain, again in matrix
notation,
\begin{widetext}
\begin{align}\label{eq:Lso3-curvature-1}
\Omega = &
\begin{bmatrix}
0 & - \frac{\sqrt{E^2-m^2}}{E m} \diff\rho \wedge \diff\theta & - \frac{\sqrt{E^2-m^2}}{E m} \sin\theta \diff\rho \wedge \diff\phi \\
\frac{\sqrt{E^2-m^2}}{E m} \diff\rho \wedge \diff\theta &  0 & -\frac{(E^2-m^2)}{ m^2}  \sin\theta \diff\theta \wedge \diff\phi \\
\frac{\sqrt{E^2-m^2}}{E m} \sin\theta \diff\rho \wedge \diff\phi & \frac{(E^2-m^2)}{ m^2}  \sin\theta \diff\theta \wedge \diff\phi  &  0
\end{bmatrix}.
\end{align}
\end{widetext}
Having constructed the connection $1$-form $\omega$ and the curvature $2$-form
$\Omega$ of the mass hyperboloid $\VC^+_{m}$ concludes the necessary geometric
ingredients we need to describe spinors on the mass hyperboloid. However, note
that at this point the picture is incomplete since it does not contain the
quantum state space, all we have is a bare basis manifold with the tangent
structure. In the next section we will finish the construction: we will add the
state space of a free spin-$1/2$ particle over the basis manifold.

\subsection{Spinors on the mass hyperboloid}
\label{sec:spinors-on-mass-hyperboloid-1}

In this section we turn to the geometric approach of the same Hilbert space
that describes a free massive spin-$1/2$ particle discussed in section
\ref{sec:qm-spins-and-twr-1}. We stress that we are not constructing a new
state space. Rather, we focus on the geometric structure that is inherent in
the same space while using the language of differential geometry to describe
how the state space of a relativistic particle arises in the geometric context.
This is common practice when working with the gauge theoretic structure of
quantum theory. We begin with an intuitive picture and then move on to a more
formal description.

Intuitively, when we think about the particle in flat spacetime, its wave
function can be represented in the position representation $\psi(x)$ or the
momentum representation $\psi(p)$. For relativistic particles, momentum space
is the mass hyperboloid $\VC^+_m$, which, as we have seen above, is a
Riemannian space with constant curvature. Then, $\psi$ is a map from $\VC^+_m$
to the Hilbert space $\hilb_{p}$ at $p\in\VC^+_m$. Technically, we can
interpret this in the language of fiber bundles, where $\VC^+_m$ is the base
manifold, $\hilb_{p}$ is the typical fiber and $\psi$ is a section of the fiber
bundle given by the union of spaces $\hilb_{p}$ over all $p\in \VC^+_m$. Common
notation in the literature for fiber bundles is $(\pi: E \rightarrow M, F)$,
where $E$ is the total space of the bundle, $M$ the base manifold, $F$ the
fiber and $\pi$ the projection from the total space to the base manifold. Hence
in the case under study we write $(\pi: E \rightarrow \VC^+_m, \hilb_{p})$. For
single particles, i.e. for spinors, the state space of the particle arises as
the space of square integrable sections of the vector bundle with a suitable
representation of the group $\SUTWO$. This state space is the space $\hilb_{m,
s}^+$ we referred to in section \ref{sec:qm-spins-and-twr-1}.

From the differential geometric perspective it is clear that one cannot
directly compare wave functions $\psi(p)$ at different points $p$ since they
belong to different spaces $\hilb_p$. In order to compare them we need to map
the wave functions into the same space, which is usually done by a non-trivial
path dependent parallel transport. The latter in general leads to non-trivial
state change as shown in Fig.~\ref{fig:boost-sequence-hyperboloid-1}.

For clarity, let us next recall how to construct the bundle for spin-$1/2$
particles, which is the spinor bundle over the three dimensional curved
manifold $\VC^+_m$ equipped with the metric $g$. In this case the typical fiber
$\hilb_p = \mathbb{C}^2$ is endowed with an action of $\SUTWO$ and spinors are
sections $\psi: \VC^+_m \to \mathbb{C}^2$.

\subsubsection{Spin group and spinors: algebra}
\label{sec:Spin}

In order to construct spinors, one starts with a Clifford algebra and then
identifies the spin group as a particular subset of the algebra. A Clifford
algebra is defined as a pair $(\mathcal{A}, \gamma)$ for a quadratic space $(V,
g)$, where $V$ is a vector space over $\RR$, $g$ a scalar product on $V$ and
$\gamma: V \rightarrow \mathcal{A}$ a linear map. The elements of the algebra
satisfy the Clifford multiplication rule
\begin{align}\label{eq:clifford-algebra-def-1}
\gamma(\mathbf{v}) \gamma(\mathbf{u}) + \gamma(\mathbf{u}) \gamma(\mathbf{v}) =
2 g(\mathbf{v}, \mathbf{u}) 1_{\mathcal{A}}
\end{align}
for all $\mathbf{u}, \mathbf{v} \in V$. In the case where $g$ has signature $p
+ q = n$, the corresponding Clifford algebra is denoted $\Cl(p, q) :=
\Cl(\RR^{p, q})$. A generic element of the algebra need not have an inverse,
hence $\Cl(p, q)$ is not a group. However, the subset of elements that do have
inverses, are normalized and consist only of products of even number of
elements, forms the spin group $\Spin(p, q)$. Specifically for our purposes $p
= 3$ and $q = 0$, then the algebra is $\Cl(3, 0)$ and the corresponding spin
group is $\Spin(3)$. It can be shown that the latter is isomorphic to $\SUTWO$.

The isomorphism provides the matrix representation of $\Spin(3)$. This way we
have arrived at \emph{spinors}: they are the elements of spaces on which the
spin group acts. In other words, spinors are real or complex column vectors
which come with the rule that specifies how they are transformed by the
elements of the Clifford algebra \cite{fecko_2006}.

Thus, Clifford algebras lead to the identification of the spin group, which in
turn gives rise to spinors.

\subsubsection{Spin and spinor bundles}
\label{sec:Spinors}

Having recalled the spin group and spinors we can construct the corresponding
bundles. How to accomplish this technically? Historically, it was not clear how
to construct spinors on Riemannian manifolds. It was only after the development
of the formalism of principal fiber bundles at the end of 1940s that spinors
could be transferred from flat spaces to Riemannian
manifolds~\cite{friedrich_2000}. Following treatments in \cite{lawson_1989,
fecko_2006}, we will construct the spinor bundle as an associated bundle of a
spin bundle. Spinor fields are defined as cross sections of the spinor bundle.

To construct a \emph{spin} bundle
$(P_{\SpinThree}\overset{\widetilde{\pi}_s}{\longrightarrow}\VC^+_m,
\SpinThree)$, we first focus on the tangent bundle of the manifold under
consideration, here the tangent bundle of the mass shell, $T \VC^+_m = E$,
which is a $6$-dimensional vector bundle over $(\VC^+_m,g)$. The bundle of
orthonormal frames $P_{\SOTHREE}(E)$ over the mass shell is a principal bundle
$(P_{\SOTHREE}(E)\overset{\widetilde{\pi}}{\longrightarrow}\VC^+_m, \SOTHREE)$,
with group $\SOTHREE$ as typical fiber.

The spin bundle is now the principle bundle in which the orthogonal group, here
$\SOTHREE$, is replaced by its double covering, here $\SpinThree \cong \SUTWO$,
such that two sheeted covering map $\xi:P_{\SpinThree}(E) \longrightarrow
P_{\SOTHREE}(E)$ exists and the following relation holds
\begin{equation}\label{fig:spin-structure-1}
	\begin{tikzcd}
		P_{\SpinThree}(E) \arrow[rr, "\xi"] \arrow[rd, "\widetilde{\pi}_s"] &       &  P_{\SOTHREE}(E)\arrow[dl, "\widetilde{\pi}"] \\
		&   \VC^+_m   &
	\end{tikzcd}
\end{equation}
where $\widetilde{\pi}$ and $\widetilde{\pi}_s$ are, respectively, projections
that define the $P_{\SOTHREE}(E)$ and $P_{\SpinThree}(E)$ bundles.

A \emph{spinor} bundle $S(E)$ of $E$ is defined as an associated vector bundle
to the spin bundle $P_{\Spin}(E)$,
\begin{align}
S(E) = P_{\Spin}(E) \times_\rho V
\end{align}
where $V$ is a vector space which carries the representation $\rho$ and action
of the group $\Spin$. For spin-$1/2$ particles over the base manifold
$(\VC^+_m,g)$, we have $\Spin = \SpinThree$, the vector space is  $V = \CC^2$
and $\rho$ is the fundamental $2 \times 2$ matrix representation of $\SpinThree
\cong \SUTWO$.

With this the construction is complete and we can regard spinor fields as
sections $\psi$ of the spinor bundle $S(E)$.

\subsubsection{The spinor connection and curvature}
\label{sec:spinor-connection-1}

Let us take stock of where we are and what needs to be accomplished next. We
have constructed the spinor bundle and defined spin states as sections of the
bundle. We have collected almost all the components needed to realize the
geometric idea of a boost as parallel transport except for one crucial
component: a connection that tells us how to transport spinors. Recall,
however, that above in \eqref{eq:local-connection-form-mass-shell-1} we
computed the connection for the tangent bundle; this describes parallel
transport of tangent vectors. It turns out that that computation was not in
vain. We can use the connection on the tangent bundle to \emph{generate} a
connection on the spinor bundle. In this section, we will explicitly compute
the spinor connection, following the discussion in references
\cite{lawson_1989, fecko_2006}.

The key idea is that the two sheeted covering $\xi$ in
(\ref{fig:spin-structure-1}) gives rise to a Lie algebra isomorphism between
the $\sutwo$ and $\sothree$ algebras, which enables to lift the $\SOTHREE$
connection to the spin connection, which in turn induces a connection on the
associated spinor bundle. Let us elaborate this in more detail.

We start by noting that we regarded the $1$-form $\omega$ as an $\SOTHREE$
connection over momentum space $\VC^+_m$. This means it maps a given vector
field $X$ on $\VC^+_m$ to $\omega(X)$, which is an element of the Lie algebra
$\sothree$, and thus $\omega$ can be written in the basis of the algebra as
\begin{align}\label{eq:so3conn}
\omega = \omega^1{}_2 E_{1}{}^{2} + \omega^1{}_3 E_{1}{}^{3} + \omega^2{}_3 E_{2}{}^{3}\, ,
\end{align}
where $E_i{}^j{}$ are the antisymmetric matrices with $-1$ at the entry $i, j$,
$1$ at the $j, i$ entry and $0$ elsewhere
\begin{align}\label{eq:skew_symmetric_elementary_matrix}
	\begin{bmatrix}
		& (i)      &         &  (j)      &         \\
		& \vdots   &         &  \vdots   &         \\
	(i)\;\cdots    & 0       & \cdots    &  -1     & \cdots  \\
	(j)\;\cdots    & 1       & \cdots    &  0      & \cdots  \\
		& \vdots   &         &  \vdots   &         \\
	\end{bmatrix},
\end{align}
which form a basis of the Lie algebra $\mathfrak{so}(3)$. The components are
given by the $1$-form components we computed in
\eqref{eq:local-connection-form-mass-shell-1},
\begin{align}
	\omega^1{}_2 = \frac{E}{m} \diff\theta\,,\quad 	\omega^1{}_3 = \sin\theta\, \frac{E}{m} \diff\phi\,,\quad \omega^2{}_3 = \cos \theta \diff\phi\,.
\end{align}

The connection $1$-form \eqref{eq:so3conn} can be elevated to a \emph{global}
connection $\widetilde{\omega}$ over the frame bundle $P_{\SOTHREE}(E)$, see an
overview of structures used in the paper in Appendix \ref{app:all-diagrams-1}.
The resulting connection contains the same information as the collection of
local connections defined on patches $\mathcal{O} \subset \VC^+_m$. The two
connections are related by a pullback $\omega = \widetilde{\sigma}^*
\widetilde{\omega}$ with a section $\widetilde{\sigma}: \VC^+_m \rightarrow
P_{\SOTHREE}(E)$ of the principal bundle. Formally, both connections are
$1$-forms taking values in the Lie algebra of the structure group of their
respective bundles. In fact, although formally $\omega$ and
$\widetilde{\omega}$ live in different spaces, the coordinate expressions of
both connections turn out to be identical, so the matrix of
$\widetilde{\omega}$ is the same as $\omega$.

The central idea in constructing the spinor connection is that once a
connection on the frame bundle $P_{\SOTHREE}(E)$ is fixed, a connection on the
spin bundle $P_{\Spin(3)}(E)$ is uniquely determined. This is because the
double covering $\xi$ induces an isomorphism $\phi$ that determines the spin
connection $\widetilde{\omega}_s$ from the frame bundle connection
$\widetilde{\omega}$, where the precise relationship is given by
\cite{fecko_2006}
\begin{align}\label{eq:relationship-omega-omega-s-1}
\xi^* \widetilde{\omega} = \phi(\widetilde{\omega}_s).
\end{align}
To obtain the spinor connection $\omega_s$ we need to pull down
$\widetilde{\omega}_s$ to the base manifold $\VC^+_m$ using a section
$\widetilde{\sigma}_s$ of $P_{\Spin(3)}$,
\begin{align}\label{eq:pulled-down-omega-s-1}
\omega_s = \widetilde{\sigma}^*_s \widetilde{\omega}_s.
\end{align}
Using also the fact that a section $\widetilde{\sigma}_s$ on the spin and a
section $\widetilde{\sigma}$ of the frame bundle are related by
$\widetilde{\sigma} = \xi \circ \widetilde{\sigma}_s$, we obtain a simple
relationship between the connections on the base manifold $\VC^+_m$,
\begin{align}\label{eq:omega-s-omega-iso-1}
\omega_s = \phi(\omega),
\end{align}
which tells us that as a practical computation for obtaining the spinor
connection $\omega_s$ from the local Levi-Civita connection $\omega$, one needs
to express the connection $\omega$ in the $\sutwo$ basis (see details in
Appendix \ref{app:spinor-connection-1}). The latter is given in terms of Pauli
matrices by
\begin{align}\label{eq:su2basis}
  J_1=-\frac{1}{2} i\sigma_1
  = -\frac{1}{2} i
  \begin{pmatrix}
    0 & 1 \\
    1 & 0
  \end{pmatrix}\,,\quad
  J_2=-\frac{1}{2} i\sigma_2
  = -\frac{1}{2} i
  \begin{pmatrix}
    0 & -i \\
    i & 0
  \end{pmatrix}\,, \quad
  J_3=-\frac{1}{2} i\sigma_3
  = -\frac{1}{2} i
  \begin{pmatrix}
    1 & 0 \\
    0 & -1
  \end{pmatrix}\,.
\end{align}
This basis satisfies
\begin{align}
    [J_i,J_k] = \sum_j \epsilon_{ikj}J^j\,.
\end{align}
The isomorphism $\phi$ between the $\sothree$ and $\sutwo$ bases is given by
\begin{align}
    E_i{}^j \leftrightarrow \epsilon_i{}^{jk} J_k\,,
\end{align}
where $\epsilon$ is the totally antisymmetric Levi-Civita tensor of $g$ in the
frame basis. Explicitly we have
\begin{align}\label{eq:isomorphism-so3-so2-algebras-1}
E_{1}{}^{2} \leftrightarrow - \frac{1}{2} i\sigma_3  = J_3  , \quad
E_{1}{}^{3} \leftrightarrow   \frac{1}{2} i\sigma_2  = -J_2 , \quad
E_{2}{}^{3} \leftrightarrow - \frac{1}{2} i\sigma_1  = J_1  .
\end{align}
Using \eqref{eq:so3conn}, the spinor connection $\omega_s$ takes the form
\begin{align}\label{eq:connSpin}
	\omega_s
	= -\frac{i}{2}  \left(
                          \omega^1{}_2 \sigma^3
                        - \omega^1{}_3 \sigma^2
                        + \omega^2{}_3 \sigma^1
                   \right)
	= - \frac{i}{2}  \left( \frac{E}{m}            \diff\theta \, \sigma^3
                        - \frac{E}{m} \sin\theta \diff\phi   \, \sigma^2
                        +  \cos\theta            \diff\phi   \, \sigma^1
                    \right)\,.
\end{align}
Similar reasoning applies to the computation of spinor curvature. Starting with
\eqref{eq:Lso3-curvature-1} which is expressed in the $\sothree$ basis, we
rewrite curvature in the $\sutwo$ basis and obtain
\begin{align}\label{eq:SinC}
\Omega_s =
  \frac{i}{2}\left(
- \frac{\sqrt{E^2-m^2}}{E m} \diff\rho \wedge \diff\theta \,         \sigma_3
+ \frac{\sqrt{E^2-m^2}}{E m} \sin\theta \diff\rho \wedge \diff\phi\, \sigma_2
- \frac{E^2-m^2}{m^2} \sin\theta \diff\theta \wedge \diff\phi \,     \sigma_1
\right).
\end{align}

These two quantities---spinor connection and curvature---represent an important
milestone. They encode the necessary geometric information for realizing the
goal we have been working for: to describe how the spin of a quantum particle
changes when it follows a path in the curved momentum space. In the next
section, we will look at a concrete example and calculate the holonomy matrix
that characterizes the state change of the spin.

\subsection{TWR as holonomy}
\label{sec:TWRHolo}

Up until now we have been claiming that in the geometric framework boosting the
particle can be understood as parallel transporting the vector which represents
the state of the particle from the initial to the final momentum. We saw above
that the simplest form of TWR occurs when the particle is boosted along a
triangular, closed path in the momentum space where the initial and the final
momenta correspond to the rest momenta of the system. This leads to the notion
of \emph{holonomy}. Holonomy is the idea that we can associate with every
closed curve $C$ a transformation matrix which maps the initial state $\psi_i$
of the system to its final state $\psi_f$ when the particle has been parallel
transported along the closed curve. The set of all such transformation matrices
forms a group which is called the \emph{holonomy group}. To determine the
transformation matrices belonging to a specific loop $C$, one needs to solve
the parallel transport equation along this loop. Formally, one can express the
transformation with help of the path ordered exponential,
\cite{Rothman:2000bz}, as
\begin{align}
    \mathrm{Hol}\left(\omega_s, C\right) = \mathcal{P}\left[\exp\left( - \int_C \omega_s \right)\right]\,.
\end{align}
This means TWR can be understood as the holonomy transformation which arises
due to the curvature of the relativistic momentum space.

By way of an example, let us calculate the holonomy matrix for a particular
boost scenario. Consider a spin-$1/2$ particle that follows a circular path $C$
in the momentum space given by
\begin{align}
\begin{split}
    C: [0,2\pi]&\to \VC^+_m\\
    \tau &\mapsto C(\tau) = (\rho(\tau), \theta(\tau), \phi(\tau)) = ( \rho_0 ,\pi/2 , \tau).
\end{split}
\end{align}
One can think of the particle that is moving with momentum of constant norm
$\rho_0 = m V/\sqrt{1-V^2}$, or with speed $V$ (see \eqref{eq:gamma}), as
undergoing infinitesimal parallel transports when it travels around the
circular trajectory. Each small boost gives rise to TWR, all of which
accumulate when the particle has completed one revolution. This is the famous
case of Thomas precession. The holonomy matrix $\mathrm{Hol}$ is a function of
connection $\omega$ and path $C$ along which the vector is parallel
transported. The holonomy of $C$ can be expressed in terms of the curvature of
the connection as
\begin{align}\label{eq:holonomy-matrix-via-curvature-1}
\mathrm{Hol}\left(\omega_s, C\right) = \exp\left( - \int_D \Omega_s \right),
\end{align}
where $D$ is the disk with boundary $C$. Using \eqref{eq:SinC} with
$\diff\theta = 0$, the holonomy integral becomes
\begin{align}
\begin{split}
	 - \int_D \Omega_s
	 &= - \int_0^{\rho_0} \int_{0}^{2\pi} \diff\rho \diff\phi\ \frac{i}{2} \frac{\rho}{\sqrt{m^2+\rho^2}m} \ \sigma_2\\
	 &= - i \sigma_2  \pi \frac{(\sqrt{m^2+\rho_0^2}-m)}{m}
	    = - i \sigma_2  \pi  \left( \frac{E(\rho_0)}{m} - 1 \right)\\
	 &= - i \sigma_2  \pi  \left( \gamma(V) - 1 \right)\,.
\end{split}
\end{align}
Introducing $\alpha = 2\pi \left( \gamma(V)- 1 \right)$ we get the following
holonomy matrix
\begin{align}\label{eq:ThoPrecFin}
	\mathrm{Hol}(\omega_s, C) = \exp\left(-i \frac{\alpha}{2} \sigma_2\right) =
	  \begin{pmatrix}
		\cos(\alpha/2) & -\sin(\alpha/2) \\
	   \sin(\alpha/2) & \cos(\alpha/2)
	\end{pmatrix}\,.
\end{align}
This $\SUTWO$ matrix acting on spin-$1/2$ particles corresponds to an
$\SOTHREE$ rotation $R_{e_2}$ by angle $\alpha$ around the $e_2 =
\frac{1}{\rho} \partial_{\theta}$ axis, see Appendix \ref{app:SU2toSO3}.

With this result we have demonstrated how to derive the TWR, and in particular
the Thomas precession of spinors, using the formalism of the differential
geometry of curved relativistic momentum space, the mass hyperboloid. Our
results coincide with the results obtained by the authors in
\cite{rhodes_relativistic_2004} for the same scenario in terms of projective
geometry. The fact that the two approaches converge on the same result
demonstrates that both reproduce the essential characteristics of the
phenomenon albeit with somewhat different means.

Let us briefly discuss the advantages offered by the two approaches. They
belong to the same family since both offer geometric conceptualization of TWR.
The approach in \cite{rhodes_relativistic_2004} provides valuable insight into
TWR by making the phenomenon easy to grasp visually as well as conceptually. It
also serves as an inspiration for the current paper.\footnote{Many facts about
the behavior of boosts can be readily grasped using projective geometry. Ref.\
\cite{rhodes_relativistic_2004} provides an excellent and accessible treatment
of TWR in the geometric setting.} On the other hand, the approach adopted here
builds on this using the language of modern differential geometry. The
advantage of the modern theory lies in that it provides a toolbox for
calculating interesting quantities in a coherent framework. For instance, it
allows to compute the holonomy matrix based on the quantities that describe the
intrinsic geometry of the mass shell manifold---the connection and curvature.
In other words, it allows to conceptualize the TWR as holonomy, associate it
with structure and relate it to the characteristics of the manifold: holonomy
forms a group which is non-trivial if the manifold has curvature.

\section{Conclusions}

We have discussed how the TWR of a spin-$1/2$ particle can be understood in a
geometric manner as the holonomy of the curved relativistic momentum space, the
mass hyperboloid. We explicitly demonstrated how to construct the spin bundle
over the mass hyperboloid and how to derive the holonomy matrix.

Since holonomies describe numerous physical effects, for example Berry's phase,
the Aharonov-Bohm effect, the Foucault pendulum, or gravitational effects on
matter wave interferometers, one sometimes groups them as being geometric or
topological, and classical or quantum holonomies \cite{lyre_2014}. For
instance, the famous case of Berry's phase is a geometric and a quantum
holonomy because it originates in the curvature of the quantum bundle. TWR, on
the other hand, is a classical and a geometric holonomy since it arises from
the curvature of the relativistic momentum space.

Having understood the TWR geometrically on Minkowski spacetime, our work paves
the way to extending this approach to curved spacetimes. To do so, we aim to
develop a geometric framework which allows to describe spinors on a curved
spacetime as sections of the spinor bundle over the non-equivalent curved
momentum spaces---the mass hyperboloids---at each point of spacetime. This
necessitates the construction of differential geometric structures that
incorporate both curved spacetimes and curved momentum spaces. A promising
candidate here is Hamilton geometry or its generalizations
\cite{Barcaroli:2015xda,Miron}. Previous approaches to describe TWR on curved
spacetimes rely on the Dirac approach to the spin-$1/2$ particle, and do not
refer to curved momentum spaces \cite{Palmer:2011bt}. Moreover, the geometric
approach to TWR allows us to study how deformed mass shells change the
predictions for the TWR. Deformed mass shells appear in deformed or doubly
special relativity (DSR) theories employed in quantum gravity phenomenology
\cite{Addazi:2021xuf}, the most studied one being the $\kappa$-Poincar\'e
framework, or in Lorentz invariance violating theories like the standard model
extension.

Thus the treatment of TWR presented here in terms of the geometry of the curved
momentum space explicitly exploits the connection between special relativistic
effects and the geometry of momentum space. It lays the foundation for future
investigations in the context of momentum spaces whose geometry differs from
the special relativistic one, and also for the addition of spacetime curvature
to the picture.

\section{Acknowledgments}

CP is funded by the excellence cluster QuantumFrontiers of the German Research
Foundation (Deutsche Forschungsgemeinschaft, DFG) under Germany's Excellence
Strategy -- EXC-2123 QuantumFrontiers -- 390837967. VP acknowledges partial
support by the Estonian Research Council (Eesti Teadusagentuur, ETAG) through
grant PSG489.

\appendix

\section{Summary of structures used in the paper}\label{app:all-diagrams-1}

The following diagram summarizes the structure and objects used in constructing
the spinor connection.

\begin{equation}\label{fig:all-diagrams-1}
\begin{tikzcd}[sep=small]
  \widetilde{\omega}_s \in \Omega^1(P_{\SpinThree}, \sutwo) \arrow[dotted, dash]{rd}  \arrow[bend right]{ddd}{\widetilde{\sigma}^*_s}   &  &   &  & \widetilde{\omega} \in \Omega^1(P_{\SOTHREE}, \sothree) \arrow[dotted,dash]{ld} \arrow[bend left]{ddd}{\widetilde{\sigma}^*} \\
  &  P_{\SpinThree}(E)   \arrow{rddd}{\widetilde{\pi}_s} \arrow{rr}{\xi}  &  & P_{\SOTHREE}(E) \arrow{lddd}[swap]{\widetilde{\pi}} & \\
  &  &  &  &  \\
\omega_s \in \Omega^1(\mathcal{O}, \sutwo) \arrow[dotted, dash]{r}    &  S(E) \arrow{rd}[swap]{\pi_s}  &  &  E \arrow{ld}{\pi}  &  \omega \in \Omega^1(\mathcal{O}, \sothree) \arrow[dotted, dash]{l} \\
  &  &  \VC^+_m  &  &
\end{tikzcd}
\end{equation}
In the diagram, $E = T\VC^+_m$ is the vector bundle---the tangent bundle of the
forward mass hyperboloid $\VC^+_m$ for particle with mass $m$. We denote
$\mathcal{O} \subset \VC_m^+$ and to avoid making the figure even more crowded,
we omit the spin section $\widetilde{\sigma}_s: \VC_m^+ \rightarrow
P_{\Spin(3)}(E)$ and the frame bundle section $\widetilde{\sigma}: \VC_m^+
\rightarrow P_{\SOTHREE}(E)$, but show how their pullbacks map connections over
principal bundles to connections for the respective associated bundles, for
instance, $\omega_s = \widetilde{\sigma}^*_s \widetilde{\omega}_s$. We also
occasionally abbreviate $P_{\SpinThree}(E) \equiv P_{\SpinThree}$ and
$P_{\SOTHREE}(E) \equiv P_{\SOTHREE}$. The dotted lines show which space a
particular connection operates on. For instance, $\omega$ belongs to the space
$\Omega^1$ of $\sothree$ algebra valued $1$-forms defined on $\mathcal{O}$, and
it defines a map on $E$. In the same vein, $\widetilde{\omega}$ is a connection
on the $\SOTHREE$ principal bundle, $\widetilde{\omega}_s$ on the $\SpinThree$
principal bundle and $\omega_s$ on the spinor bundle $S(E)$.

\section{The spinor connection}
\label{app:spinor-connection-1}

Given the (local) Levi-Civita connection $1$-form $\omega$ on $\VC^+_m$, we can
compute the (local) spinor connection $\omega_s$ on $\VC^+_m$. We follow the
discussions in \cite{lawson_1989, fecko_2006}. Using the notation of
\eqref{fig:all-diagrams-1}, we denote connections on the corresponding
principal bundles with $\widetilde{\phantom{\omega}}$, i.e.\ the global
connection $\widetilde{\omega}$ on $P_{\SOTHREE}(E)$ and the spin connection
$\widetilde{\omega}_s$ on $P_{\Spin}(E)$. The relationship between the two
connections is given by
\begin{align}\label{eq:relationship-omega-omega-s-2}
\xi^* \widetilde{\omega} = \phi(\widetilde{\omega}_s).
\end{align}
In order to obtain the spinor connection $\omega_s$ we need to pull down
$\widetilde{\omega}_s$ to the base manifold $\VC^*_m$ using a section
$\widetilde{\sigma}_s$ of the spin bundle $P_{\Spin(3)}(E)$,
\begin{align}\label{eq:pulled-down-omega-s-2}
\omega_s = \widetilde{\sigma}^*_s \widetilde{\omega}_s.
\end{align}
We can now check that the pulled down sections $\widetilde{\sigma}^*
\widetilde{\omega}$ and $\widetilde{\sigma}^*_s \widetilde{\omega}_s$ are
related by the isomorphism $\phi$. Using the fact that a section
$\widetilde{\sigma}_s$ on the spin and a section $\widetilde{\sigma}$ of the
frame bundle are related by $\widetilde{\sigma} = \xi \circ
\widetilde{\sigma}_s$, we compute
\begin{align}\label{eq:isomorphism-between-pulldowns-1}
\widetilde{\sigma}^* \widetilde{\omega} &= (\xi \circ \widetilde{\sigma}_s)^* \widetilde{\omega} \\
&= \widetilde{\sigma}_s^* (\xi^* \, \widetilde{\omega}) \\
&= \widetilde{\sigma}_s^* \left(\phi (\widetilde{\omega}_s) \right) \\
&= \widetilde{\sigma}_s^* \left( \widetilde{\omega}_s^A \, \phi(E_A) \right) \\
&= \left( \widetilde{\sigma}_s^* \widetilde{\omega}_s^A \right) \phi(E_A) \\
&= \phi \left( (\widetilde{\sigma}_s^* \widetilde{\omega}_s^A) E_A \right) \\
&= \phi \left( \widetilde{\sigma}_s^* \widetilde{\omega}_s \right).
\end{align}
Since the pulldown of the Levi-Civita connection $\omega = \widetilde{\sigma}^*
\widetilde{\omega}$, we obtain the isomorphism between connections on the base
manifold $\VC^+_m$,
\begin{align}\label{eq:omega-s-omega-iso-2}
\omega = \phi(\omega_s).
\end{align}

\section{Identifying $\SUTWO$ and $\SOTHREE$}\label{app:SU2toSO3}

Below equation \eqref{eq:ThoPrecFin} we claimed that the obtained $\SUTWO$
holonomy transformation corresponds to an $\SOTHREE$ rotation around the $e_2$
axis. Here we quickly recall the relation between $\SUTWO$ elements and
$\SOTHREE$ rotations for completeness and for a self contained discussion.

Consider $(M,h)$ being a $3$-dimensional Riemannian manifold. Let $Z = Z^a e_a
\in T_xM \sim \mathbb{R}^3$, where $e_a$ is a orthonormal basis of $T_xM$,
i.e.\ $h(e_a,e_b)=\delta_{ab}$. Let $\hat Z = \frac{Z}{h(Z,Z)}$, be the
normalization of $Z$ and define the $\mathfrak{su}(2)$ representation
$\mathbf{Z}$ of $Z$ as $\mathbf{Z}=Z^aJ_a$ (the $J_a$ are defined in
\eqref{eq:su2basis}). Then
\begin{align}
	U(\mathbf{Z}) = \exp\left( \mathbf{Z}\right) = \mathbf{1} \cos\left( \frac{\sqrt{h(Z,Z)}}{2} \right) - 2 \left( \frac{\mathbf{Z}}{\sqrt{h(Z,Z)}} \right)\sin\left(\frac{\sqrt{h(Z,Z)}}{2}\right)\,,
\end{align}
is a representation of an element $U(\mathbf{Z})$ of $\SUTWO$, generated by the
$\mathfrak{su}(2)$ element $\mathbf{Z}$, on $\mathbb{C}^2$, parameterized by the
components $Z^a$ of the vector $Z$. It is easy to check that $\det
(U(\mathbf{Z})) = 1$. Call $\sqrt{h(Z,Z)} = \varphi$, then the $\SUTWO$
elements become
\begin{align}
 U(\mathbf{Z}) = \exp \left(\mathbf{Z}\right)
 = \mathbf{1} \cos \left( \frac{\varphi}{2} \right) - 2 \mathbf{\hat Z} \sin\left( \frac{\varphi}{2} \right)
 = \mathbf{1} \cos \left( \frac{\varphi}{2} \right) + i \hat Z^a \sigma_a \sin\left( \frac{\varphi}{2} \right)\,.
\end{align}
They corresponds to a $\SOTHREE$ element $R_Z$ which represents a rotation
around the $Z$-axis of an angle $\varphi$ in the following way.

Let $X^a e_a$ and $Y=Y^a e_a$ be vectors in $T_xM$ expanded in a orthonormal
basis of $T_x M$. Then, they can be identified with the elements
$\mathbf{X}=X^a J_a$ and $\mathbf{Y} = Y^a J_a$ in $\mathfrak{su}(2)$. The
scalar product is encoded as
\begin{align}
	h(X,Y) = -2 \textrm{Tr}\left( \mathbf{X}\ \mathbf{Y} \right)\,.
\end{align}
Define
\begin{align}
	R_Z(X) = R_{U(\mathbf{Z})}(\mathbf{X}) = U(\mathbf{Z})\ \mathbf{X}\ U(\mathbf{Z})^{-1}\,
\end{align}
which is a map from $T_x M$ to $T_x M$. It preserves the scalar product
\begin{align}
	h(X,Y) = - 2\textrm{Tr}\left(  \mathbf{X}\ \mathbf{Y}\right) = - 2\textrm{Tr}\left( R_{U(\mathbf{Z})}(\mathbf{X}) R_{U(\mathbf{Z})}(\mathbf{Y})\right)  = h(R_{Z}(X),R_{Z}(Y))
\end{align}
Thus $R_Z(X)$ are the rotations of $X$ around the $Z$-axis by an angle $\varphi
= h(Z,Z)$ expressed in terms of $\SUTWO$ elements $U(\mathbf{Z})$. It is clear
that $U(Z)$ and $-U(Z)$ generate the same rotations, which makes visible the
fact that $\SUTWO$ is the double covering of $\SOTHREE$.

\bibliography{paper_2021_wigner-holonomy}

\begin{thebibliography}{30}%
\makeatletter
\providecommand \@ifxundefined [1]{%
 \@ifx{#1\undefined}
}%
\providecommand \@ifnum [1]{%
 \ifnum #1\expandafter \@firstoftwo
 \else \expandafter \@secondoftwo
 \fi
}%
\providecommand \@ifx [1]{%
 \ifx #1\expandafter \@firstoftwo
 \else \expandafter \@secondoftwo
 \fi
}%
\providecommand \natexlab [1]{#1}%
\providecommand \enquote  [1]{``#1''}%
\providecommand \bibnamefont  [1]{#1}%
\providecommand \bibfnamefont [1]{#1}%
\providecommand \citenamefont [1]{#1}%
\providecommand \href@noop [0]{\@secondoftwo}%
\providecommand \href [0]{\begingroup \@sanitize@url \@href}%
\providecommand \@href[1]{\@@startlink{#1}\@@href}%
\providecommand \@@href[1]{\endgroup#1\@@endlink}%
\providecommand \@sanitize@url [0]{\catcode `\\12\catcode `\$12\catcode
  `\&12\catcode `\#12\catcode `\^12\catcode `\_12\catcode `\%12\relax}%
\providecommand \@@startlink[1]{}%
\providecommand \@@endlink[0]{}%
\providecommand \url  [0]{\begingroup\@sanitize@url \@url }%
\providecommand \@url [1]{\endgroup\@href {#1}{\urlprefix }}%
\providecommand \urlprefix  [0]{URL }%
\providecommand \Eprint [0]{\href }%
\providecommand \doibase [0]{http://dx.doi.org/}%
\providecommand \selectlanguage [0]{\@gobble}%
\providecommand \bibinfo  [0]{\@secondoftwo}%
\providecommand \bibfield  [0]{\@secondoftwo}%
\providecommand \translation [1]{[#1]}%
\providecommand \BibitemOpen [0]{}%
\providecommand \bibitemStop [0]{}%
\providecommand \bibitemNoStop [0]{.\EOS\space}%
\providecommand \EOS [0]{\spacefactor3000\relax}%
\providecommand \BibitemShut  [1]{\csname bibitem#1\endcsname}%
\let\auto@bib@innerbib\@empty
%</preamble>
\bibitem [{\citenamefont {Thomas}(1926)}]{ThomasOriginal}%
  \BibitemOpen
  \bibfield  {author} {\bibinfo {author} {\bibfnamefont {L.~H.}\ \bibnamefont
  {Thomas}},\ }\href {\doibase 10.1038/117514a0} {\bibfield  {journal}
  {\bibinfo  {journal} {Nature}\ }\textbf {\bibinfo {volume} {117}},\ \bibinfo
  {pages} {514} (\bibinfo {year} {1926})}\BibitemShut {NoStop}%
\bibitem [{\citenamefont {Wigner}(1939)}]{WignerOrig}%
  \BibitemOpen
  \bibfield  {author} {\bibinfo {author} {\bibfnamefont {E.}~\bibnamefont
  {Wigner}},\ }\href {http://www.jstor.org/stable/1968551} {\bibfield
  {journal} {\bibinfo  {journal} {Annals of Mathematics}\ }\textbf {\bibinfo
  {volume} {40}},\ \bibinfo {pages} {149} (\bibinfo {year} {1939})}\BibitemShut
  {NoStop}%
\bibitem [{\citenamefont {Malykin}(2006)}]{Malykin_2006}%
  \BibitemOpen
  \bibfield  {author} {\bibinfo {author} {\bibfnamefont {G.~B.}\ \bibnamefont
  {Malykin}},\ }\href {\doibase 10.1070/PU2006v049n08ABEH005870} {\bibfield
  {journal} {\bibinfo  {journal} {Physics-Uspekhi}\ }\textbf {\bibinfo {volume}
  {49}},\ \bibinfo {pages} {837} (\bibinfo {year} {2006})}\BibitemShut
  {NoStop}%
\bibitem [{\citenamefont {O'Donnell}\ and\ \citenamefont
  {Visser}(2011)}]{ODonnell:2011ivw}%
  \BibitemOpen
  \bibfield  {author} {\bibinfo {author} {\bibfnamefont {K.}~\bibnamefont
  {O'Donnell}}\ and\ \bibinfo {author} {\bibfnamefont {M.}~\bibnamefont
  {Visser}},\ }\href {\doibase 10.1088/0143-0807/32/4/016} {\bibfield
  {journal} {\bibinfo  {journal} {Eur. J. Phys.}\ }\textbf {\bibinfo {volume}
  {32}},\ \bibinfo {pages} {1033} (\bibinfo {year} {2011})},\ \Eprint
  {http://arxiv.org/abs/1102.2001} {arXiv:1102.2001 [gr-qc]} \BibitemShut
  {NoStop}%
\bibitem [{\citenamefont {Kholmetskii}\ and\ \citenamefont
  {Yarman}(2020)}]{Kholmetskii_2020}%
  \BibitemOpen
  \bibfield  {author} {\bibinfo {author} {\bibfnamefont {A.~L.}\ \bibnamefont
  {Kholmetskii}}\ and\ \bibinfo {author} {\bibfnamefont {T.}~\bibnamefont
  {Yarman}},\ }\href {\doibase 10.1088/1361-6404/ab8e27} {\bibfield  {journal}
  {\bibinfo  {journal} {European Journal of Physics}\ }\textbf {\bibinfo
  {volume} {41}},\ \bibinfo {pages} {055601} (\bibinfo {year}
  {2020})}\BibitemShut {NoStop}%
\bibitem [{\citenamefont {Rhodes}\ and\ \citenamefont
  {Semon}(2004)}]{rhodes_relativistic_2004}%
  \BibitemOpen
  \bibfield  {author} {\bibinfo {author} {\bibfnamefont {J.~A.}\ \bibnamefont
  {Rhodes}}\ and\ \bibinfo {author} {\bibfnamefont {M.~D.}\ \bibnamefont
  {Semon}},\ }\href {\doibase 10.1119/1.1652040} {\bibfield  {journal}
  {\bibinfo  {journal} {Am. J. Phys.}\ }\textbf {\bibinfo {volume} {72}},\
  \bibinfo {pages} {943} (\bibinfo {year} {2004})}\BibitemShut {NoStop}%
\bibitem [{\citenamefont {Aravind}(1997)}]{aravind_1997}%
  \BibitemOpen
  \bibfield  {author} {\bibinfo {author} {\bibfnamefont {P.~K.}\ \bibnamefont
  {Aravind}},\ }\href@noop {} {\bibfield  {journal} {\bibinfo  {journal}
  {American Journal of Physics}\ }\textbf {\bibinfo {volume} {65}},\ \bibinfo
  {pages} {634} (\bibinfo {year} {1997})}\BibitemShut {NoStop}%
\bibitem [{\citenamefont {Criado}\ and\ \citenamefont
  {Alamo}(2001)}]{criado_2001}%
  \BibitemOpen
  \bibfield  {author} {\bibinfo {author} {\bibfnamefont {C.}~\bibnamefont
  {Criado}}\ and\ \bibinfo {author} {\bibfnamefont {N.}~\bibnamefont {Alamo}},\
  }\href@noop {} {\bibfield  {journal} {\bibinfo  {journal} {American Journal
  of Physics}\ }\textbf {\bibinfo {volume} {69}},\ \bibinfo {pages} {306}
  (\bibinfo {year} {2001})}\BibitemShut {NoStop}%
\bibitem [{\citenamefont {Lyre}(2014)}]{lyre_2014}%
  \BibitemOpen
  \bibfield  {author} {\bibinfo {author} {\bibfnamefont {H.}~\bibnamefont
  {Lyre}},\ }\href {\doibase 10.1016/j.shpsb.2014.08.013} {\bibfield  {journal}
  {\bibinfo  {journal} {Studies in History and Philosophy of Science Part B:
  Studies in History and Philosophy of Modern Physics}\ }\textbf {\bibinfo
  {volume} {48}},\ \bibinfo {pages} {45} (\bibinfo {year} {2014})}\BibitemShut
  {NoStop}%
\bibitem [{\citenamefont {Criado}\ and\ \citenamefont
  {Alamo}(2009)}]{CRIADO2009923}%
  \BibitemOpen
  \bibfield  {author} {\bibinfo {author} {\bibfnamefont {C.}~\bibnamefont
  {Criado}}\ and\ \bibinfo {author} {\bibfnamefont {N.}~\bibnamefont {Alamo}},\
  }\href {\doibase 10.1016/j.ijnonlinmec.2009.06.008} {\bibfield  {journal}
  {\bibinfo  {journal} {International Journal of Non-Linear Mechanics}\
  }\textbf {\bibinfo {volume} {44}},\ \bibinfo {pages} {923} (\bibinfo {year}
  {2009})}\BibitemShut {NoStop}%
\bibitem [{\citenamefont {Peres}\ \emph {et~al.}(2002)\citenamefont {Peres},
  \citenamefont {Scudo},\ and\ \citenamefont {Terno}}]{peres_quantum_2002}%
  \BibitemOpen
  \bibfield  {author} {\bibinfo {author} {\bibfnamefont {A.}~\bibnamefont
  {Peres}}, \bibinfo {author} {\bibfnamefont {P.~F.}\ \bibnamefont {Scudo}}, \
  and\ \bibinfo {author} {\bibfnamefont {D.~R.}\ \bibnamefont {Terno}},\ }\href
  {\doibase 10.1103/PhysRevLett.88.230402} {\bibfield  {journal} {\bibinfo
  {journal} {Physical Review Letters}\ }\textbf {\bibinfo {volume} {88}},\
  \bibinfo {pages} {230402} (\bibinfo {year} {2002})}\BibitemShut {NoStop}%
\bibitem [{\citenamefont {Gingrich}\ and\ \citenamefont
  {Adami}(2002)}]{gingrich_quantum_2002}%
  \BibitemOpen
  \bibfield  {author} {\bibinfo {author} {\bibfnamefont {R.~M.}\ \bibnamefont
  {Gingrich}}\ and\ \bibinfo {author} {\bibfnamefont {C.}~\bibnamefont
  {Adami}},\ }\href {\doibase 10.1103/PhysRevLett.89.270402} {\bibfield
  {journal} {\bibinfo  {journal} {Physical Review Letters}\ }\textbf {\bibinfo
  {volume} {89}},\ \bibinfo {pages} {270402} (\bibinfo {year}
  {2002})}\BibitemShut {NoStop}%
\bibitem [{\citenamefont {Caban}\ \emph {et~al.}(2009)\citenamefont {Caban},
  \citenamefont {Rembieli{\'n}ski},\ and\ \citenamefont
  {W{\l}odarczyk}}]{caban_2009_strange}%
  \BibitemOpen
  \bibfield  {author} {\bibinfo {author} {\bibfnamefont {P.}~\bibnamefont
  {Caban}}, \bibinfo {author} {\bibfnamefont {J.}~\bibnamefont
  {Rembieli{\'n}ski}}, \ and\ \bibinfo {author} {\bibfnamefont
  {M.}~\bibnamefont {W{\l}odarczyk}},\ }\href
  {https://journals.aps.org/pra/abstract/10.1103/PhysRevA.79.014102} {\bibfield
   {journal} {\bibinfo  {journal} {Physical Review A}\ }\textbf {\bibinfo
  {volume} {79}},\ \bibinfo {pages} {014102} (\bibinfo {year}
  {2009})}\BibitemShut {NoStop}%
\bibitem [{\citenamefont {Friis}\ \emph {et~al.}(2010)\citenamefont {Friis},
  \citenamefont {Bertlmann}, \citenamefont {Huber},\ and\ \citenamefont
  {Hiesmayr}}]{friis_2010_relativistic}%
  \BibitemOpen
  \bibfield  {author} {\bibinfo {author} {\bibfnamefont {N.}~\bibnamefont
  {Friis}}, \bibinfo {author} {\bibfnamefont {R.~A.}\ \bibnamefont
  {Bertlmann}}, \bibinfo {author} {\bibfnamefont {M.}~\bibnamefont {Huber}}, \
  and\ \bibinfo {author} {\bibfnamefont {B.~C.}\ \bibnamefont {Hiesmayr}},\
  }\href {https://journals.aps.org/pra/abstract/10.1103/PhysRevA.81.042114}
  {\bibfield  {journal} {\bibinfo  {journal} {Physical Review A}\ }\textbf
  {\bibinfo {volume} {81}},\ \bibinfo {pages} {042114} (\bibinfo {year}
  {2010})}\BibitemShut {NoStop}%
\bibitem [{\citenamefont {Palge}\ and\ \citenamefont
  {Dunningham}(2015)}]{palge_2015_werner}%
  \BibitemOpen
  \bibfield  {author} {\bibinfo {author} {\bibfnamefont {V.}~\bibnamefont
  {Palge}}\ and\ \bibinfo {author} {\bibfnamefont {J.}~\bibnamefont
  {Dunningham}},\ }\href {\doibase 10.1016/j.aop.2015.09.028} {\bibfield
  {journal} {\bibinfo  {journal} {Annals of Physics}\ }\textbf {\bibinfo
  {volume} {363}},\ \bibinfo {pages} {275} (\bibinfo {year}
  {2015})}\BibitemShut {NoStop}%
\bibitem [{\citenamefont {Palge}\ \emph {et~al.}(2019)\citenamefont {Palge},
  \citenamefont {Dunningham}, \citenamefont {Groote},\ and\ \citenamefont
  {Liivat}}]{palge_2019_maps}%
  \BibitemOpen
  \bibfield  {author} {\bibinfo {author} {\bibfnamefont {V.}~\bibnamefont
  {Palge}}, \bibinfo {author} {\bibfnamefont {J.}~\bibnamefont {Dunningham}},
  \bibinfo {author} {\bibfnamefont {S.}~\bibnamefont {Groote}}, \ and\ \bibinfo
  {author} {\bibfnamefont {H.}~\bibnamefont {Liivat}},\ }\href
  {https://www.worldscientific.com/doi/abs/10.1142/S123016121950001X}
  {\bibfield  {journal} {\bibinfo  {journal} {Open Systems \& Information
  Dynamics}\ }\textbf {\bibinfo {volume} {26}},\ \bibinfo {pages} {1950001}
  (\bibinfo {year} {2019})}\BibitemShut {NoStop}%
\bibitem [{\citenamefont {Barr}\ \emph {et~al.}(2023)\citenamefont {Barr},
  \citenamefont {Caban},\ and\ \citenamefont
  {Rembieli{\'n}ski}}]{barr_2023_bell}%
  \BibitemOpen
  \bibfield  {author} {\bibinfo {author} {\bibfnamefont {A.~J.}\ \bibnamefont
  {Barr}}, \bibinfo {author} {\bibfnamefont {P.}~\bibnamefont {Caban}}, \ and\
  \bibinfo {author} {\bibfnamefont {J.}~\bibnamefont {Rembieli{\'n}ski}},\
  }\href {https://doi.org/10.22331/q-2023-07-27-1070} {\bibfield  {journal}
  {\bibinfo  {journal} {Quantum}\ }\textbf {\bibinfo {volume} {7}},\ \bibinfo
  {pages} {1070} (\bibinfo {year} {2023})}\BibitemShut {NoStop}%
\bibitem [{\citenamefont {Polyzou}\ \emph {et~al.}(2012)\citenamefont
  {Polyzou}, \citenamefont {Gl{\"o}ckle},\ and\ \citenamefont
  {Wita{\l}a}}]{polyzou_spin_2012}%
  \BibitemOpen
  \bibfield  {author} {\bibinfo {author} {\bibfnamefont {W.~N.}\ \bibnamefont
  {Polyzou}}, \bibinfo {author} {\bibfnamefont {W.}~\bibnamefont
  {Gl{\"o}ckle}}, \ and\ \bibinfo {author} {\bibfnamefont {H.}~\bibnamefont
  {Wita{\l}a}},\ }\href {\doibase 10.1007/s00601-012-0526-8} {\bibfield
  {journal} {\bibinfo  {journal} {Few-Body Systems}\ }\textbf {\bibinfo
  {volume} {54}},\ \bibinfo {pages} {1667} (\bibinfo {year}
  {2012})}\BibitemShut {NoStop}%
\bibitem [{\citenamefont {Bogolubov}\ \emph {et~al.}(1975)\citenamefont
  {Bogolubov}, \citenamefont {Logunov},\ and\ \citenamefont
  {Todorov}}]{bogolubov_introduction_1975}%
  \BibitemOpen
  \bibfield  {author} {\bibinfo {author} {\bibfnamefont {N.~N.}\ \bibnamefont
  {Bogolubov}}, \bibinfo {author} {\bibfnamefont {A.~A.}\ \bibnamefont
  {Logunov}}, \ and\ \bibinfo {author} {\bibfnamefont {I.~T.}\ \bibnamefont
  {Todorov}},\ }\href@noop {} {\emph {\bibinfo {title} {Introduction to
  {Axiomatic} {Quantum} {Field} {Theory}}}}\ (\bibinfo  {publisher}
  {W.A.Benjamin},\ \bibinfo {year} {1975})\BibitemShut {NoStop}%
\bibitem [{\citenamefont {Sexl}\ and\ \citenamefont
  {Urbantke}(2001)}]{sexl_relativity_2001}%
  \BibitemOpen
  \bibfield  {author} {\bibinfo {author} {\bibfnamefont {R.~U.}\ \bibnamefont
  {Sexl}}\ and\ \bibinfo {author} {\bibfnamefont {H.~K.}\ \bibnamefont
  {Urbantke}},\ }\href@noop {} {\emph {\bibinfo {title} {Relativity, {Groups},
  {Particles}: {Special} {Relativity} and {Relativistic} {Symmetry} in {Field}
  and {Particle} {Physics}}}},\ \bibinfo {edition} {rev. ed.}\ ed.\ (\bibinfo
  {publisher} {Springer},\ \bibinfo {address} {New York},\ \bibinfo {year}
  {2001})\BibitemShut {NoStop}%
\bibitem [{\citenamefont {Caban}\ \emph {et~al.}(2013)\citenamefont {Caban},
  \citenamefont {Rembieli{\'n}ski},\ and\ \citenamefont
  {W{\l}odarczyk}}]{caban_spin_2013}%
  \BibitemOpen
  \bibfield  {author} {\bibinfo {author} {\bibfnamefont {P.}~\bibnamefont
  {Caban}}, \bibinfo {author} {\bibfnamefont {J.}~\bibnamefont
  {Rembieli{\'n}ski}}, \ and\ \bibinfo {author} {\bibfnamefont
  {M.}~\bibnamefont {W{\l}odarczyk}},\ }\href
  {http://journals.aps.org/pra/abstract/10.1103/PhysRevA.88.022119} {\bibfield
  {journal} {\bibinfo  {journal} {Physical Review A}\ }\textbf {\bibinfo
  {volume} {88}},\ \bibinfo {pages} {022119} (\bibinfo {year}
  {2013})}\BibitemShut {NoStop}%
\bibitem [{\citenamefont {Halpern}(1968)}]{halpern_special_1968}%
  \BibitemOpen
  \bibfield  {author} {\bibinfo {author} {\bibfnamefont {F.~R.}\ \bibnamefont
  {Halpern}},\ }\href@noop {} {\emph {\bibinfo {title} {Special {Relativity}
  and {Quantum} {Mechanics}}}}\ (\bibinfo  {publisher} {Prentice-Hall},\
  \bibinfo {year} {1968})\BibitemShut {NoStop}%
\bibitem [{\citenamefont {Fecko}(2006)}]{fecko_2006}%
  \BibitemOpen
  \bibfield  {author} {\bibinfo {author} {\bibfnamefont {M.}~\bibnamefont
  {Fecko}},\ }\href@noop {} {\emph {\bibinfo {title} {Differential geometry and
  Lie groups for physicists}}}\ (\bibinfo  {publisher} {Cambridge University
  Press},\ \bibinfo {year} {2006})\BibitemShut {NoStop}%
\bibitem [{\citenamefont {Friedrich}(2000)}]{friedrich_2000}%
  \BibitemOpen
  \bibfield  {author} {\bibinfo {author} {\bibfnamefont {T.}~\bibnamefont
  {Friedrich}},\ }\href@noop {} {\emph {\bibinfo {title} {Dirac Operators in
  Riemannian Geometry}}}\ (\bibinfo  {publisher} {American Mathematical
  Society},\ \bibinfo {address} {Providence, Rhode Island},\ \bibinfo {year}
  {2000})\BibitemShut {NoStop}%
\bibitem [{\citenamefont {Lawson}\ and\ \citenamefont
  {Michelsohn}(1989)}]{lawson_1989}%
  \BibitemOpen
  \bibfield  {author} {\bibinfo {author} {\bibfnamefont {H.~B.}\ \bibnamefont
  {Lawson}}\ and\ \bibinfo {author} {\bibfnamefont {M.-L.}\ \bibnamefont
  {Michelsohn}},\ }\href@noop {} {\emph {\bibinfo {title} {Spin geometry}}}\
  (\bibinfo  {publisher} {Princeton University Press},\ \bibinfo {year}
  {1989})\BibitemShut {NoStop}%
\bibitem [{\citenamefont {Rothman}\ \emph {et~al.}(2001)\citenamefont
  {Rothman}, \citenamefont {Ellis},\ and\ \citenamefont
  {Murugan}}]{Rothman:2000bz}%
  \BibitemOpen
  \bibfield  {author} {\bibinfo {author} {\bibfnamefont {T.}~\bibnamefont
  {Rothman}}, \bibinfo {author} {\bibfnamefont {G.~F.~R.}\ \bibnamefont
  {Ellis}}, \ and\ \bibinfo {author} {\bibfnamefont {J.}~\bibnamefont
  {Murugan}},\ }\href {\doibase 10.1088/0264-9381/18/7/306} {\bibfield
  {journal} {\bibinfo  {journal} {Class. Quant. Grav.}\ }\textbf {\bibinfo
  {volume} {18}},\ \bibinfo {pages} {1217} (\bibinfo {year} {2001})},\ \Eprint
  {http://arxiv.org/abs/gr-qc/0008070} {arXiv:gr-qc/0008070 [gr-gc]}
  \BibitemShut {NoStop}%
\bibitem [{\citenamefont {Barcaroli}\ \emph {et~al.}(2015)\citenamefont
  {Barcaroli}, \citenamefont {Brunkhorst}, \citenamefont {Gubitosi},
  \citenamefont {Loret},\ and\ \citenamefont {Pfeifer}}]{Barcaroli:2015xda}%
  \BibitemOpen
  \bibfield  {author} {\bibinfo {author} {\bibfnamefont {L.}~\bibnamefont
  {Barcaroli}}, \bibinfo {author} {\bibfnamefont {L.~K.}\ \bibnamefont
  {Brunkhorst}}, \bibinfo {author} {\bibfnamefont {G.}~\bibnamefont
  {Gubitosi}}, \bibinfo {author} {\bibfnamefont {N.}~\bibnamefont {Loret}}, \
  and\ \bibinfo {author} {\bibfnamefont {C.}~\bibnamefont {Pfeifer}},\ }\href
  {\doibase 10.1103/PhysRevD.92.084053} {\bibfield  {journal} {\bibinfo
  {journal} {Phys. Rev. D}\ }\textbf {\bibinfo {volume} {92}},\ \bibinfo
  {pages} {084053} (\bibinfo {year} {2015})},\ \Eprint
  {http://arxiv.org/abs/1507.00922} {arXiv:1507.00922 [gr-qc]} \BibitemShut
  {NoStop}%
\bibitem [{\citenamefont {Miron}\ \emph {et~al.}(2001)\citenamefont {Miron},
  \citenamefont {Dragos}, \citenamefont {Shimada},\ and\ \citenamefont
  {Sabau}}]{Miron}%
  \BibitemOpen
  \bibfield  {author} {\bibinfo {author} {\bibfnamefont {R.}~\bibnamefont
  {Miron}}, \bibinfo {author} {\bibfnamefont {H.}~\bibnamefont {Dragos}},
  \bibinfo {author} {\bibfnamefont {H.}~\bibnamefont {Shimada}}, \ and\
  \bibinfo {author} {\bibfnamefont {S.}~\bibnamefont {Sabau}},\ }\href@noop {}
  {\emph {\bibinfo {title} {The geometry of Hamilton and Lagrange Spaces}}}\
  (\bibinfo  {publisher} {Kluwer Academic},\ \bibinfo {year}
  {2001})\BibitemShut {NoStop}%
\bibitem [{\citenamefont {Palmer}\ \emph {et~al.}(2012)\citenamefont {Palmer},
  \citenamefont {Takahashi},\ and\ \citenamefont {Westman}}]{Palmer:2011bt}%
  \BibitemOpen
  \bibfield  {author} {\bibinfo {author} {\bibfnamefont {M.~C.}\ \bibnamefont
  {Palmer}}, \bibinfo {author} {\bibfnamefont {M.}~\bibnamefont {Takahashi}}, \
  and\ \bibinfo {author} {\bibfnamefont {H.~F.}\ \bibnamefont {Westman}},\
  }\href {\doibase 10.1016/j.aop.2011.10.009} {\bibfield  {journal} {\bibinfo
  {journal} {Annals Phys.}\ }\textbf {\bibinfo {volume} {327}},\ \bibinfo
  {pages} {1078} (\bibinfo {year} {2012})},\ \Eprint
  {http://arxiv.org/abs/1108.3896} {arXiv:1108.3896 [quant-ph]} \BibitemShut
  {NoStop}%
\bibitem [{\citenamefont {Addazi}\ \emph {et~al.}(2022)\citenamefont {Addazi}
  \emph {et~al.}}]{Addazi:2021xuf}%
  \BibitemOpen
  \bibfield  {author} {\bibinfo {author} {\bibfnamefont {A.}~\bibnamefont
  {Addazi}} \emph {et~al.},\ }\href {\doibase 10.1016/j.ppnp.2022.103948}
  {\bibfield  {journal} {\bibinfo  {journal} {Prog. Part. Nucl. Phys.}\
  }\textbf {\bibinfo {volume} {125}},\ \bibinfo {pages} {103948} (\bibinfo
  {year} {2022})},\ \Eprint {http://arxiv.org/abs/2111.05659} {arXiv:2111.05659
  [hep-ph]} \BibitemShut {NoStop}%
\end{thebibliography}%

\end{document}